# Detection of two new RRATs at 111 MHz

S.V. Logvinenko[1], S.A. Tyul'bashev[1], V.M. Malofeev[1]

[1]Lebedev Physics Institute, Pushchino radio astronomy observatory;

142290, Russia, Moscow reg., Pushchino, PRAO FIAN

**Abstract**

Two new rotating transients were detected in the 2020 observations carried out on the LPA LPI radio telescope. The dispersion measures of the found transients are DM=21 and 35 pc/cm$^3$, the pulse half-widths are $W_e$=18 and 35 ms for J1550+09 and J2047+13, respectively. The upper estimate of the period RRAT J2047+13 P=2.925s was obtained. The study shows the existence of rotating transients whose pulses appear less frequently than one pulse per 10 hours of observations.

**Introduction**

Rotating radio transients (RRAT) were discovered in 2006 as sporadically appearing dispersed pulses, the frequency of which varied from several minutes to several hours [1]. The dispersion measure of the detected pulses or the distance to the objects indicated their galactic nature, and the periods and derivatives of periods that could be estimated for many RRAT corresponded to ordinary second pulsars. Thus, it is very likely that RRAT is a special kind of pulsar. The total number of detected RRAT is not large. About 90 RRAT were detected in the decimeter range (see ATNF catalog https://www.atnf.csiro.au/people/pulsar/psrcat / ). Approximately 30 RRAT were detected in the meter range in observations on a Large Phased Array of the Lebedev Physical Institute (LPA LPI) [2,3].

The nature of RRAT is still not clear. In [4], it is assumed that RRAT are ordinary pulsars with strong pulses. If this is the case, then after removing the detected strong dispersed pulses from the raw data, a weak periodic radiation should appear. In [5] it is stated that RRAT are pulsars with extreme nullings. The nulling phenomenon consists in the fact that from time to time pulsars have a pulse skip [6]. For some pulsars, nulling can be complex. For example, in the pulsar J0810+37, discovered in the survey at the LPA LPI [7], the proportion of missing pulses varies from 38% to 74% regularly throughout the year of observations [8]. That is, for the pulsar J0810+37, the number of observed pulses may be less than the number of missed pulses. It is believed that the greater the characteristic age of the pulsar, the greater the value of nulling [9]. If the hypothesis of extreme nullings [5] is confirmed, then RRAT are old pulsars. There are also more exotic hypotheses. For example, it is assumed that the RRAT pulse is observed when a neutron star collides with matter falling on it in the form of asteroids, or due to the interaction of the pulsar magnetosphere with the matter of the accretion disk located around the star [10,11]. It is possible that different hypotheses are true for different transients.

Monitoring observations have been going on at Pushchino Radio Astronomy Observatory (PRAO) for several years, which can be used to search for pulsars and rotating transients. In this paper, the search for transients using a new processing program is considered.

## Observations and results

Daily observations in monitoring mode are carried out on the LPA LPI radio telescope in 32-channel mode with a channel width of 78 kHz and a readout time of 12.5 ms. The large effective area of the radio telescope, which is approximately 45,000 sq.m., provides high fluctuation sensitivity, which makes it possible to search for RRAT. Details about the radio telescope and observation programs can be found in [12].

Since observations are conducted simultaneously in 96 spatial beams at declinations from -7º to +42º, a specially designed and manufactured in PRAO recorder is used for permanent data recording. Structurally, it consists of an industrial computer (PC), into which the required number of basic modules is installed (see Fig.1). When installing 6 base modules, 48 signals or beams of a radio telescope are processed simultaneously. The expansion of the number of simultaneously recorded signals can be carried out by installing additional PCs equipped with the necessary number of digital receiver modules (up to 8 modules).

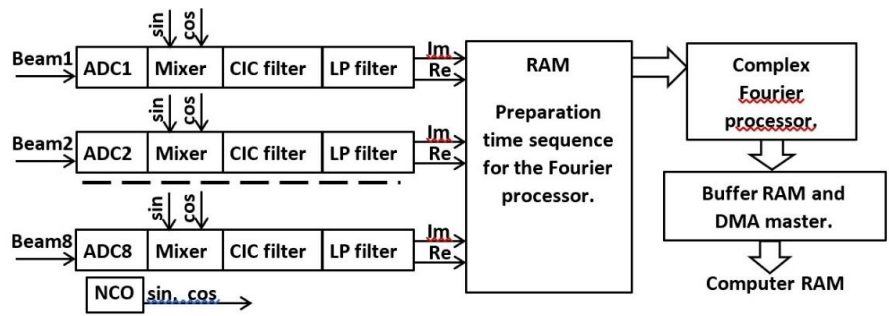

Fig.1 Basic recorder module. ADC is an analog-to-digital converter. CIC filter – Cascaded Integrator-Comb (CIC) Filter. KPDP is a direct access channel to memory. LPF is low-pass filter. RAM is a random access memory device.

The signal from each beam is digitized at a frequency of 230 MHz (direct digitization of the signal) using a digital video converter with an operating band of 109-111.5 MHz, and transferred to the low frequency region. Next, a spectral analysis of the signal is performed with accumulation up to a given time interval of reading.

The module is made in the PCI standard, data exchange is carried out over the universal PCI bus, which allows the installation of the module in computers equipped with a PCI bus of both the +5V standard and the +3.3V. The frequency of data exchange over the PCI bus is 33 MHz, 32 bits, which meets the requirements for data exchange speed for this task.

To digitize the signal, a 12-bit, 2-channel ADC from Texas Instruments ADS62P29 is used with a maximum digitization frequency of 250 MHz. To reduce the influence of interference and reduce the inter-channel mutual penetration of signals, a separate system of stabilization and filtering of supply voltages is used. In addition, the topology of the printed circuit board provides means to reduce the impact of interference penetrating the signal over common buses. Data transmission from the ADC to the PLD (Programmable Logic Device) is carried out via differential lines according to the LVDS standard with double data transfer rate. To reduce the number of components on the printed circuit board from the PLD side, built-in PLD means of matching data transmission lines are used.

Data from 8 ADCs of one module is transmitted via differential communication lines. In addition, the PCI interface is implemented on the module and means of intermodule data exchange are provided. These features dictated the need to use PLDs with a large number of input/output (I/O) pins and with the possibility of implementing signal processing with a frequency of 230 MHz. Therefore, the module uses Altera's EP3SL780C3 PLD – the youngest PLD model from the

StratixIII family with a 3-speed gradation. This chip is equipped with 50,000 equivalent logic elements and has 480 user I/O elements. PLD resources make it possible to implement 8 independent video converters on one chip, filtering of high-frequency and low-frequency signals, spectral analysis and processing of 8 independent data streams. The use of sequentially connected CIC filters and LF filters allows you to digitally suppress unnecessary signal and interference located outside the operating band of the radio telescope by at least 80 dB. Such suppression, combined with the use of analog filters of the telescope's receiving system, makes it possible to realize the necessary sensitivity of the radio telescope.

The drivers necessary for the organization of module management by means of the operating system were developed using the Microsoft Driver Development Kit (DDK). The developed drivers allow you to organize simultaneous operation of up to 8 modules on one PC, including direct memory access mode.

Previously, 33 rotating transients were detected during the processing of semi-annual monitoring data from July-December 2015 [2,3]. In these observations, the values of dispersion measures (DM) up to 100 pc/cm$^3$ were sorted during processing. As is known, the DM reflects the total electron density along the line of sight. The greater the total electron density, the greater the DM at the same distance to the object under study. If, when testing the next DM, pulses with a signal-to-noise ratio (S/N) greater than 7 appeared, the coordinates of the detected object and data were recorded in the catalog, allowing to generate an image with an profile and a dynamic spectrum of the transient. In particular, strong pulses can be detected with different S/N in a whole set of different DM. In addition, some industrial interference with the formal addition of frequency channels may show DM typical of pulsars. When processing half a year of monitoring observations, about one million objects were found that formally meet the criteria of the new RRAT. It turned out that most of the detected transients are pulses of known pulsars. For a strong pulsar, more than a thousand pulses can be detected for each observation session. However, even after the exclusion of known pulsars, the visual verification of candidates took a long time.

A new observation processing program has been developed to search for RRAT. In this program, when checking the DM for a given day and for a given direction, there is only one candidate showing the highest S/N ratio. Known pulsars and RRAT, strong scintillating radio sources, calibration signals recorded 6 times a day show known and false transients. Such objects are included manually or automatically in specially created directories. This makes it possible to exclude them from the final directories with candidates for transients, which has reduced the number of candidates being checked tenfold.

After processing for several days in February 2020, 2 new rotating transients c S/N>7 were detected at declinations -7°<δ<+21° (J1550+09: $\alpha_{2000}$=15$^h$50$^m$47$^s$±90$^s$; $\delta_{2000}$=9°43'±15'; DM=21±1.5 pc/cm$^3$; $W_e$=18 ms; J2047+12: $\alpha_{2000}$=20$^h$47$^m$45$^s$±90$^s$; $\delta_{2000}$=12°59'±15'; DM=36±2 pc/cm$^3$; $W_e$=35 ms). With the new processing of the available archival data, RRAT J1549+09 was detected 4 times in the interval of 4 years, and J2047+12 was detected 7 times. The dynamic spectra and profiles of the RRAT pulses are shown in Fig.2. For the transient J2047+12, two pulses were detected on one of the days at a distance of 2.925s. The actual period of the pulsar can be a multiple of a number of times less.

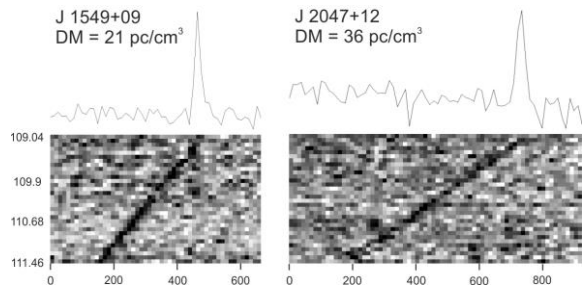

Fig.2 The time in milliseconds is postponed along the horizontal axis. On the vertical axis is the frequency of observations in MHz. The dark line inside the dynamic spectrum is the signal from the detected transient. The intensity of radiation within the dynamic spectrum is shown in shades of gray. The darker the color, the stronger the recorded signal. Pulse profiles are shown above the spectra, representing the integral sum of the pulse in all frequency channels, combined taking into account the dependence of the signal delay in the interstellar medium on the frequency of observations. The degree of slope of the line reflects the magnitude of the DM. The greater the slope, the greater the DM.

**Discussion of the results and conclusion**

The galactic coordinates RRAT J1550+09 (l=19°19'; b=+44°21') and RRAT J2047+12 (l=58°58'; b=-18°36') indicate that both transients found are far beyond the plane of the Galaxy. Distance estimates based on the NE2001 model [13], based on the DM of the objects found and the distribution of electrons in the Galaxy, show R=0.95 kpc (J1550+09) and R=2.2 kpc (J2047+12). These distances are typical for pulsars.

According to the work on the search for rotating transients, the time between sporadically appearing pulses can be in the range from minutes to hours [1,14]. When searching for RRAT on a semi-annual interval [3], the equivalent observation time of each point in the sky at declinations from -7° to +42° was approximately 10 hours. Therefore, it was expected that in the studied area all strong (S/N>7) RRAT available for LPA LPI had already been detected. After detecting new RRATs, the results of an early transient search conducted on semi-annual data were rechecked. It is confirmed that both transients are missing in the early processing data. The average observation time before the appearance of the pulse from J1550+09 turned out to be 20 hours, and for J2047+12 – 11 hours. Thus, in the meter range there are RRAT, which have one pulse for 10 or more hours. Long series of observations and the use of the programs described above make it possible to detect such rarely flashing rotating radio transients.

**Bibliography**

[1] M. McLaughlin, A. Lyne, D. Lorimer, et al., Transient radio bursts from rotating neutron stars, Nature (2006) **439**, pp.817-820 (DOI:10.1038/nature04440)

[2] S.A. Tyul'bashev, V.S. Tyul'bashev, V.M. Malofeev, et al., Detection of five new RRATs at 111 MHz, Astronomy Reports (2018) **62**, pp. 63-71 (DOI: 10.1134/S1063772918010079)

[3] S.A. Tyul'bashev, V.S. Tyul'bashev, V.M. Malofeev, Detection of 25 new rotating radio transients at 111 MHz, A&A (2018a) **618**, A70 (DOI: 10.1051/0004-6361/201833102)

[4] P. Weltevrede, B.W. Stappers, J.M. Rankin, G.A.E. Wright, Is pulsar B0656+14 a very nearby rotating radio transient, ApJ (2006) **645**, pp. L149-L152 (DOI: 10.1086/506346)

[5] B. Zhang, J. Gil, J. Dyks, On the origins of part-time radio pulsars, MNRAS (2007) **374**, pp. 1103-1107 (DOI: 10.1111/j.1365-2966.2006.11226.x)

[6] D. Backer, Pulsar nulling phenomena, Nature, (1970) **228,** pp.42-43 (DOI:10.1038/228042a0)

[7] S.A. Tyul'bashev, V.S. Tyul'bashev, M.A. Kitaeva, et al., Search for and detection of pulsars in monitoring observations at 111 MHz, Astronomy Reports (2017), **61**, pp. 848-858 (doi:10.1134/S1063772917100109)

[8] D.A. Teplykh, V.M. Malofeev, Nulling Phenomenon of the New Radio Pulsar J0810+37 at a Frequency of 111 MHz, Bulletin of the Lebedev Physics Institute (2019), **46**, pp. 380-382 (doi:10.3103/S1068335619120030)

[9] V. Gajjar, On the absence of pulses from pulsars, PhD thesis, 2014, National Centre for Radio Astrophysics, Tata Institute of Fundamental Research, Pune University (https://arxiv.org/abs/1706.05407)

[10] X.-D. Li, On the nature of part-time radio pulsars, ApJ (2006) **646**, pp. L139-L142 (DOI: 10.1086/506962)

[11] Q. Luo, D. Melrose, Pulsar radiation belts and transient radio emission, MNRAS (2007) **378**, pp.1481-1490 (DOI: 10.1111/j.1365-2966.2007.11889.x)

[12] S.A. Tyul'bashev, V.S. Tyul'bashev, V.V. Oreshko, S.V. Logvinenko, Detection of new pulsars at 111 MHz, Astronomy Reports (2016) **60**, pp. 220-232 (DOI: 10.1134/S1063772916020128)

[13] J. Cordes, T. Lazio, NE2001. II. Using radio propagation data to construct a model for the Galactic distribution of free electrons, ArXiv e-prints., 2002 [arXiv:astro-ph/0301598]

[14] S. Burke-Spolaor, M. Bailes, The millisecond radio sky: transients from a blind single-pulse search, MNRAS (2010) **402**, pp.855-866 (DOI: 10.1111/j.1365-2966.2009.15965.x)